\documentclass{llncs}
\usepackage{makeidx}  
\usepackage{graphicx}        
\usepackage{fancyvrb}
\usepackage{fixltx2e}
\usepackage{amsfonts}
\usepackage{alltt}
\usepackage[polish, german, english]{babel}
\begin{document}
\title{Optimizing data collection for object tracking in wireless sensor networks}
\titlerunning{Optimizing data collection for object tracking in wireless sensor networks}  
%
\selectlanguage{polish}
\author{Bart"lomiej P"laczek\inst{1} \and Marcin Berna"s\inst{2}}
\authorrunning{Bart"lomiej P"laczek and Marcin Berna"s} 
\institute{Faculty of Transport, Silesian University of Technology,\\Krasi"nskiego 8, 40-019 Katowice, Poland\\
\email{placzek.bartlomiej@gmail.com} \and Institute of Computer Science, University of Silesia, B"edzi"nska 39, 41-200 Sosnowiec, Poland\\
\email{marcin.bernas@gmail.com}
}

\maketitle              

\selectlanguage{english}

\begin{abstract}
In this paper some modifications are proposed to optimize an algorithm of object tracking in wireless sensor network (WSN). The task under consideration is to control movement of a mobile sink, which has to reach a target in the shortest possible time. Utilization of the WSN resources is optimized by transferring only selected data readings (target locations) to the mobile sink. Simulations were performed to evaluate the proposed modifications against state-of-the-art methods. The obtained results show that the presented tracking algorithm allows for substantial reduction of data collection costs with no significant increase in the amount of time that it takes to catch the target.\selectlanguage{polish}\footnote{Preprint of: P"laczek B., Berna"s M.: Optimizing Data Collection for Object Tracking in Wireless Sensor Networks. Communications in Computer and Information Science, vol. 370, pp. 485-494 (2013). The final publication is available at www.springerlink.com}
\end{abstract}
\selectlanguage{english}
\section{Introduction}
The task of object tracking in wireless sensor network (WSN) is to detect a moving object (target), localize it and report its location to the sink. Usually, it is assumed that the actual location of the object has to be determined continuously with a predetermined precision. The tracking capabilities of WSNs have been used in many applications, such as battlefield monitoring, wildlife habitat monitoring, intruder detection, and traffic control \cite{bplaczek:bib1,bplaczek:bib2}. In this paper we consider an application of the object tracking WSN for target chasing. It means that the location information is delivered to a mobile sink, which has to follow and catch the target. Thus, the objective of target chasing is to control the movement of a mobile sink which has to reach the target in the shortest possible time.

The design of object tracking applications over WSNs raise great challenges due to the bandwidth-limited communication medium, energy constraints, data congestion and transmission delays. These issues are particularly important when dealing with the target chasing task, which require reliable real-time data delivery \cite{bplaczek:bib3}. In order to ensure efficient execution of this task, the data collection procedures for WSN should be optimized, taking into account the above-mentioned limitations. Continuous data collection scheme is not suitable for developing the target chasing applications, because periodical transmissions of the target location to the sink would drain sensors energy rapidly. Therefore, the target chasing task requires dedicated data collection methods to ensure the amount of transmitted data is as low as possible. 

In this paper the existing WSN-based target chasing methods are discussed and some modifications are proposed to optimize the use of sensor nodes by transferring only selected data readings (target locations) to the mobile sink. Simulation experiments were performed to evaluate the proposed modifications against state-of-the-art methods. The experimental results show that the presented tracking algorithm allows for substantial reduction in the amount of transmitted data with no significant negative effect on performance of the target chasing application.

The remainder of this paper is organized as follows. Related works are reviewed in Section 2. Section 3 presents a detailed description of the considered target chasing problem. WSN-based target tracking algorithms and data collection strategies are presented in Section 4. Section 5 contains results of the experiments on the target chasing in WSN. Finally, in Section 6, conclusion is given.
%
\section{Related works}
In the literature a number of methods have been introduced for wireless sensor networks that enable the tracking of moving objects. A comprehensive and detailed review of these approaches can be found in \cite{bplaczek:bib1,bplaczek:bib4}. The majority of the available methods aim at delivering the real-time information about trajectory of a tracked object to a single sink. However, in most approaches the sink is assumed to be stationary and only few publications deal with the problem of chasing the target by a mobile sink.

A basic formulation of the target chasing problem postulates that the target performs a simple random walk in a two dimensional grid, moving to one of the four adjacent grid points with equal probability every time step \cite{bplaczek:bib5}. The chasing strategy presented in \cite{bplaczek:bib5} was designed for the case of static sensors able to detect the target, with no communication between them. A static sensor can deliver the information about when the target was detected to the mobile sink only if the sink is located on the same grid point as the sensor.

More realistic model of the WSN was used by Tsai et al. \cite{bplaczek:bib6} to develop the dy-namical object tracking protocol (DOT). This protocol allows the sensor network to assist in detecting the target and collecting the target’s trajectory information. The trajectory information is stored by an intermediate (beacon) node, which guides the sink to chase the target. A similar method is the target tracking with monitor and backup sensors \cite{bplaczek:bib7}, which additionally take into account the effect of a target’s variable velocity and direction.

Complex scenarios with multiple targets and multiple pursuers are also analysed in the literature \cite{bplaczek:bib3,bplaczek:bib8}. For such scenarios a centralized coordination of the pursuers has to be performed by a control module, i.e. by a base station or one of the pursuers. This task requires both communication among pursuers and high computational resources. The tracking algorithms discussed in this paper can contribute to the complex scenarios by optimizing the data transmission between the control module and particular pursuers.

The proposed approach extends the DOT protocol by providing heuristic rules to reduce the amount of data transmitted in WSN during target chasing. Moreover, in order to decrease the number of activated sensor nodes the introduced data collection strategies adopt the prediction-based tracking method \cite{bplaczek:bib9}. According to this method a prediction model is used, which anticipates the future location of the target so only the sensor nodes expected to detect the target are periodically activated.

\section{Problem formulation}
The considered target chasing problem deals with controlling movement of a mobile sink, which has to catch the single target in the shortest possible time. The target moves in a closed area, which is divided into square segments of equal dimensions. Discrete Cartesian coordinates $(x, y) \in  \mathbb{N}^2$ are used to identify the segments. For each segment there is a static sensor node deployed that can detect presence of the target in this particular segment. Communication range of each sensor node covers the four nearest neighbouring segments $(x+1, y), (x-1, y), (x, y+1)$, and $(x, y-1)$ . At a given time, only the selected sensor nodes are in active state to track the target and other nodes are put into sleep state to save energy consumption.

The chasing procedure is executed in discrete time steps. At each time step both the target and the sink move in one of the four directions: north, west, south or east. Their velocities (in segments per time step) are determined as parameters of the simulation. Target changes its movement direction randomly. The probability that the target moves to an adjacent segment depends on the direction. Moving direction of the sink is decided on the basis of information delivered from WSN.

Main objective in target chasing is to minimize time-to-catch, i.e., the number of time steps in which the sink reaches the moving target. However, due to the limited energy resources also the minimization of data collection costs (data transmission and sensing cost) has to be taken into consideration. The data collection costs are measured by: number of data transfers to the sink, hop counts, and active times of sensor nodes. An obvious trade-off exists between the time-to-catch minimization and the data collection costs minimization. In this study heuristic rules are proposed that enable considerable reduction of the data collection costs with no significant increase of time-to-catch. 

\section{Tracking algorithms}
In this study, three object tracking algorithms are compared in application to the target chasing problem. The first two algorithms presented below were developed on the basis of the methods available in literature, i.e. the prediction-based tracking and the dynamical object tracking. The last part of this section describes the proposed algorithm, which utilizes heuristic rules to select data readings that have to be transmitted.

\subsection{Prediction-based tracking}
According to Algorithm 1 (Fig. 1), which uses the prediction-based tracking method, the target location is discovered and reported to the sink at each time step. The sink moves toward segment $(x_T, y_T)$, where the target is detected. It means that movement direction is selected which minimizes distance between the sink location $(x_S, y_S)$ and the target location $(x_T, y_T)$. Because the sink can move in one of the four directions (N, W, S, E), the city-block metric was used to determine the distance $D$ between segments:
\begin{equation}
   D[(x_1, y_1), (x_2, y_2)] = \vert x_1-x_2 \vert +  \vert y_1-y_2 \vert.
\end{equation}

Thus, the selected segment $(x_S^*, y_S^*)$ into which the sink will move, has to fulfil the following condition: 
\begin{equation}
   (x_S^*, y_S^*) \in M \land D[(x_S^*, y_S^*),(x_T, y_T)] = \min_{(x, y) \in M} \lbrace D[(x, y),(x_T, y_T)]  \rbrace,
\end{equation}
where $M = \lbrace ( x_S+v_s \cdot \Delta T, y_S), ( x_S-v_s \cdot \Delta T, y_S), ( x_S, y_S+v_s \cdot \Delta T), ( x_S, y_S-v_s \cdot \Delta T)$ is the set of segments that can be selected, $v_S$ denotes value of the sink velocity, and $\Delta T = 1$ time step. 
%
%
\begin{figure}
\begin{alltt}
1 \hspace{0.3cm} set node(x\textsubscript{C},y\textsubscript{C}) to be target node
2 \hspace{0.3cm} repeat
3 \hspace{0.6cm} at target node do
4 \hspace{0.9cm} determine P
5 \hspace{0.9cm} collect data from each node (x,y):(x,y) \(\in\) P
6 \hspace{0.9cm} determine (x\textsubscript{T},y\textsubscript{T})
7 \hspace{0.9cm} if (x\textsubscript{T},y\textsubscript{T}) changed then
8 \hspace{1.2cm} communicate (x\textsubscript{T},y\textsubscript{T}) to the sink
9 \hspace{1.2cm} set node(x\textsubscript{T},y\textsubscript{T}) to be target node
10 \hspace{0.6cm} at sink do 
11 \hspace{0.9cm} move toward (x\textsubscript{T},y\textsubscript{T})
12 \hspace{0.3cm} until (x\textsubscript{S},y\textsubscript{S})=(x\textsubscript{T},y\textsubscript{T})
\end{alltt}
\caption{Pseudocode of Algorithm 1}
\end{figure}
Prediction of the possible target locations is based on a simple model of the target movement, which is consistent with the above assumptions on available directions and predetermined maximum velocity. Let us denote maximum value of the target velocity by $v_T$. If for previous time step $(t-1)$ the target was detected in segment $(x_T, y_T)$, then at time $t$ there is a set $P$ of possible target locations: 
\begin{equation}
   P = \lbrace (x, y): D[(x, y), (x_T, y_T)] \leq v_T \cdot \Delta T \rbrace,
\end{equation}
where $\Delta T = 1$ time step.

Sensor nodes for all possible target locations $(x, y) \in P$ are activated, and the discovered target location is transmitted to the sink. The transmission is suppressed if the target location is the same as at the previous time step. At the beginning of the tracking procedure a central segment $(x_C, y_C)$ of the monitored area is assumed to be a hypothetical target location. 

An important feature of the above algorithm is that the collected information about target trajectory has the maximum available precision. Moreover, the information is delivered to the sink with the highest attainable frequency (at each time step of the tracking procedure).

\subsection{Dynamical object tracking}
Pseudocode of Algorithm 2 is shown in Fig. 2. This algorithm is based on the tracking method which was proposed for the DOT protocol \cite{bplaczek:bib6}. In this algorithm the location of target is discovered at each time step using the same approach as in Algorithm 1. 

The target location $(x_T, y_T)$ is determined and stored at the intermediate node. Sink moves toward location of the intermediate node $(x_I, y_I)$. A new intermediate node is set if the sink enters segment $(x_I, y_I)$ or a predetermined time $\tau$ passes since the last update. The update means that the sensor node, which currently detects the target in segment $(x_T, y_T)$, becomes new intermediate node and its location is communicated to the sink.
%
%
\begin{figure}
\begin{alltt}
1 \hspace{0.3cm} in case of Algorithm 2
2 \hspace{0.6cm} condition:= \(\Delta\)t=\(\tau\) or (x\textsubscript{S},y\textsubscript{S})=(x\textsubscript{I},y\textsubscript{I}) 
3 \hspace{0.3cm} in case of Algorithm 3
4 \hspace{0.6cm} condition:= d\textsubscript{IT}/d\textsubscript{SI}>\(\alpha\) or \(\Delta\)t/(d\textsubscript{ST}/v\textsubscript{S})>\(\beta\) or (x\textsubscript{S},y\textsubscript{S})=(x\textsubscript{I},y\textsubscript{I})
5 \hspace{0.3cm} set node(x\textsubscript{C},y\textsubscript{C}) to be intermediate node
6 \hspace{0.3cm} \(\Delta\)t:=0
7 \hspace{0.3cm} repeat
8 \hspace{0.6cm} at intermediate node do
9 \hspace{0.9cm} determine P
10 \hspace{0.9cm} collect data from each node(x,y):(x,y)\(\in\)P
11 \hspace{0.9cm} determine (x\textsubscript{T},y\textsubscript{T})
12 \hspace{0.9cm} if (x\textsubscript{T},y\textsubscript{T})<>(x\textsubscript{I},y\textsubscript{I}) then
13 \hspace{1.2cm} \(\Delta\)t:=\(\Delta\)t+1
14 \hspace{1.2cm} if condition then
15 \hspace{1.5cm} \(\Delta\)t:=0
16 \hspace{1.5cm} set node(x\textsubscript{T},y\textsubscript{T}) to be intermediate node	
17 \hspace{1.5cm} communicate new (x\textsubscript{I},y\textsubscript{I}) to the sink
18 \hspace{0.6cm} at sink do 
19 \hspace{1.2cm} move toward (x\textsubscript{I},y\textsubscript{I})
20 \hspace{0.3cm} until (x\textsubscript{S},y\textsubscript{S})=(x\textsubscript{T},y\textsubscript{T})
\end{alltt}
\caption{Pseudocode of Algorithms 2 and 3}
\end{figure}

By using the intermediate node, the cost of data transmission in WSN is reduced because the data transfers to sink are executed less frequently. Data readings from the activated sensor nodes are collected by the intermediate node, which is closer to the segments $(x, y) \in P$ than the sink. Therefore, a lower number of hops is required to complete the data transmission.

\subsection{Application of heuristic rules}
Algorithm 3 implements heuristic rules that are proposed to improve performance of the object tracking task in terms of data collection costs. The major difference between Algorithm 2 and Algorithm 3 lies in the condition, which determines when the intermediate node has to be updated. The following symbols are used to formulate this condition (Fig. 2, line 4 of the pseudocode): $d_{ST}$ – distance between sink and target, $d_{IT}$ – distance between intermediate node and target, $d_{SI}$ – distance between sink and intermediate node.

The heuristic rules were motivated by an observation that the sink does not need the precise information about target location to chase the target effectively when the distance to target is large. The closer to target, the higher precision of the localization has to be obtained. This fact is illustrated by examples in Fig. 3. Locations of the target and the sink are indicated by ‘T’ and ‘S’. The numbers describe distance to target from segments into which the sink can move during the analysed time step. Optimal moves are shown by arrows. It was assumed that $v_S = 2$ and $v_T = 0$ for these examples. Required precision of the target localization corresponds to the shaded regions. It should be noted here that the sink moves toward intermediate node. The shaded regions indicate segments where the intermediate node can be located to ensure that the sink selects the optimal movement direction. It can be seen from this illustration that for a greater distance between sink and target there is a larger area, which includes allowable locations of the intermediate node. 
%
%
\begin{figure}
\centering
\includegraphics [width=11.6cm] {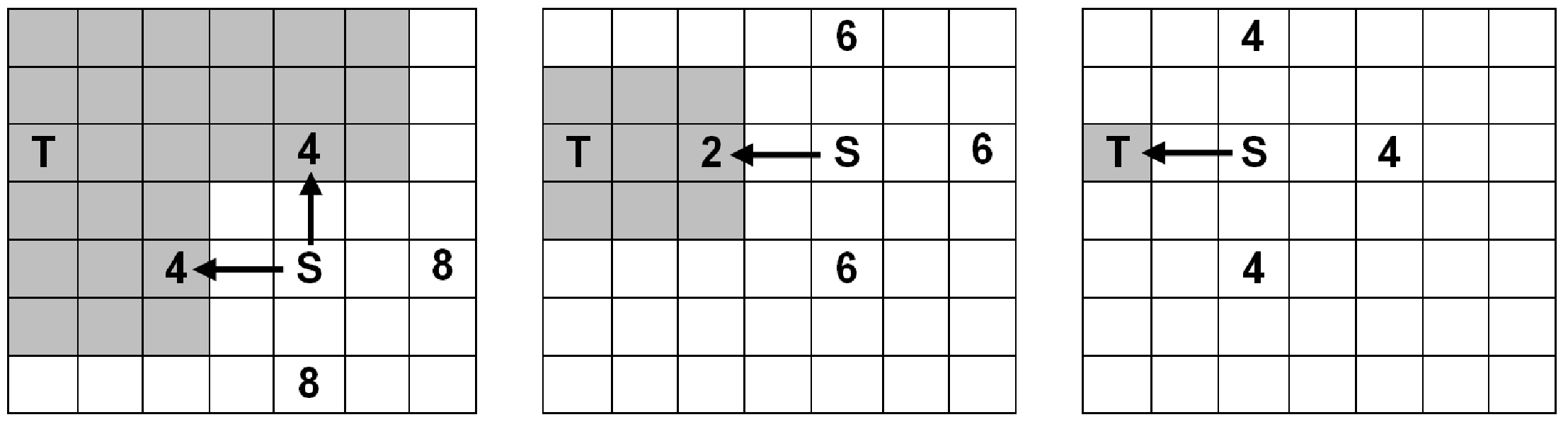}
\caption{Required precision of the target location information}
\end{figure}

On the basis of the above insights two heuristic rules were introduced. The first rule says that the update of intermediate node is necessary if the distance between intermediate node and target is relatively high in comparison to the distance between sink and intermediate node. This rule was translated into the elementary condition  $d_{IT} / d_{SI} > \alpha$. According to the second rule, the update of intermediate node has to be executed if the time elapsed since the last update is relatively long in comparison to the time in which the sink could reach the current target location. The second rule can be written as a formula:  $\Delta t / (d_{ST} / v_{S}) > \beta$. Values of the parameters $\alpha$ and $\beta$ were chosen experimentally.

\section{Experimental results}
Simulation experiments were performed to compare effectiveness of the three tracking algorithms presented in Section 4. The comparison was made with respect to data collection costs and tracking performance. Hop counts, active times of sensor nodes and numbers of data transfers to sink were analyzed to evaluate the cost of data collection in WSN. The tracking performance was measured as time-to-catch, i.e., the time in which the sink reaches the moving target. Both the active time and the time-to-catch are measured in time steps of the control procedure. Hop counts were determined assuming that the shortest path is used for each data transfer.

In the experiments, it was assumed that target velocity $v_T$ equals 3 and sink velocity $v_S$ equals 4. It should be noted that the velocities are expressed in segments per time step. The monitored area is a square of 200 x 200 segments. Thus, the number of sensor nodes equals 40 000. The results presented below were averaged for 100 simulation runs. Each simulation run starts with the same locations of sink and target. During simulation, random trajectory of the target is generated. The simulation stops when target is caught by the sink. Experiments were performed using simulation software that was developed for this research.
%
%
\begin{figure}
\centering
\includegraphics [width=12cm] {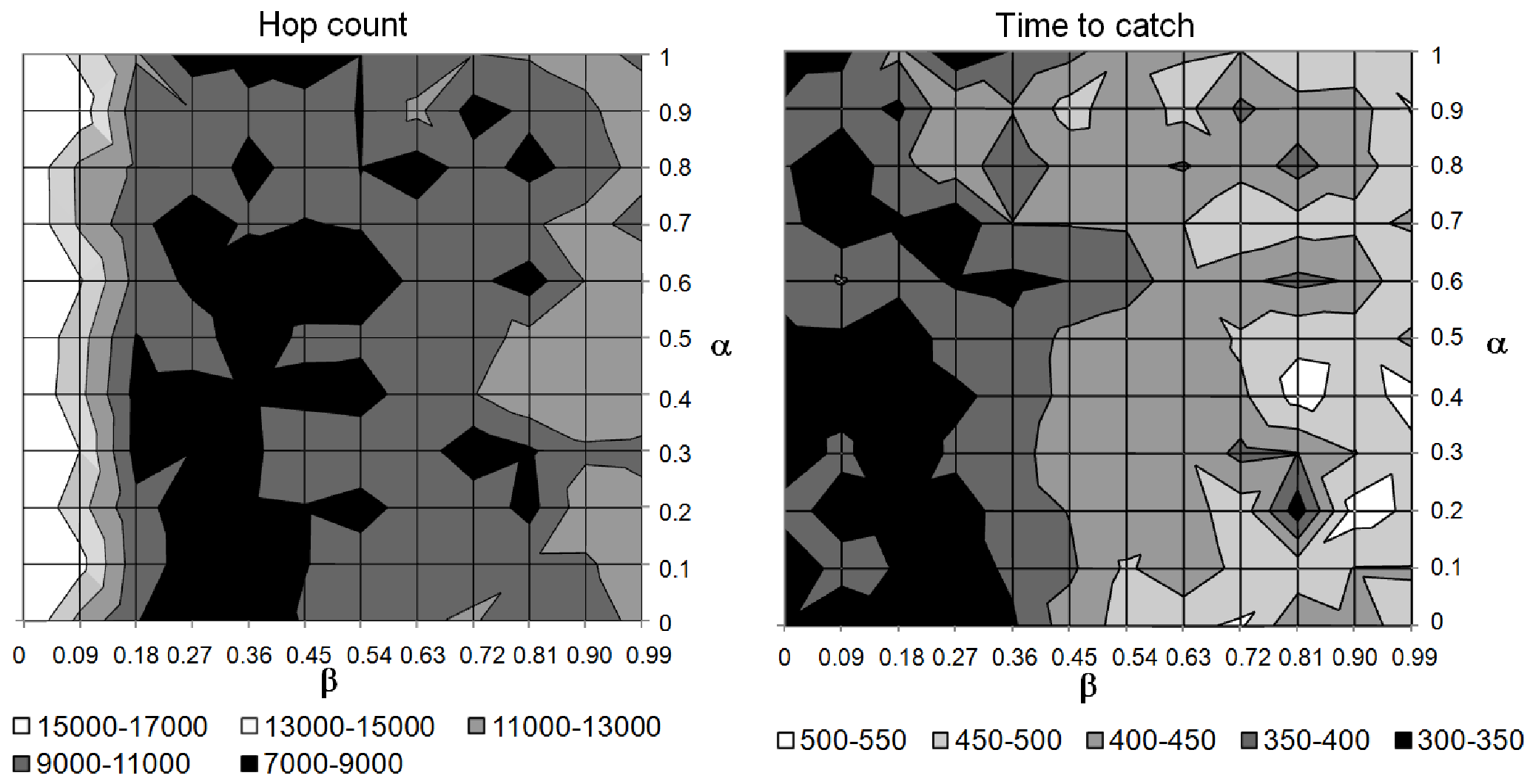}
\caption{Impact of parameters $\alpha$ and $\beta$ on hop count and time-to-catch for Algorithm 3}
\end{figure}

Initial experiments were carried out in order to examine influence of the parameters $\alpha$ and $\beta$ on effectiveness of Algorithm 3. The results in Fig. 4 illustrate the effect of these parameters on hop count as well as on time-to-catch. Black colour in these charts corresponds with low level of the analyzed quantities. A similar analysis was conducted for total active time of sensor nodes and number of data transfers to sink. Based on the results, the optimal values of parameters were determined: $\alpha = 0.20$ and $\beta = 0.25$. In case of Algorithm 2 the preliminary experiments have shown that the best results can be obtained for $\tau = 6$. The above settings were used for all the simulations reported in this section. 

In Fig. 5 simulation results are compared for the three examined algorithms. From these results it is apparent that the proposed algorithm (Algorithm 3) provides short time-to-catch values with low data collection cost. In comparison, Algorithm 1 involves a much higher number of data transfers and hops. Algorithm 2 needs significantly longer time to reach the moving target than the other considered algorithms. As it could be expected, the shortest time-to-catch was obtained for Algorithm 1, in which the extracted information about target locations has the highest available precision. However, for Algorithms 1 and 3 the difference of time-to-catch values is negligible.
%
%
\begin{figure}
\centering
\includegraphics [width=9.3cm] {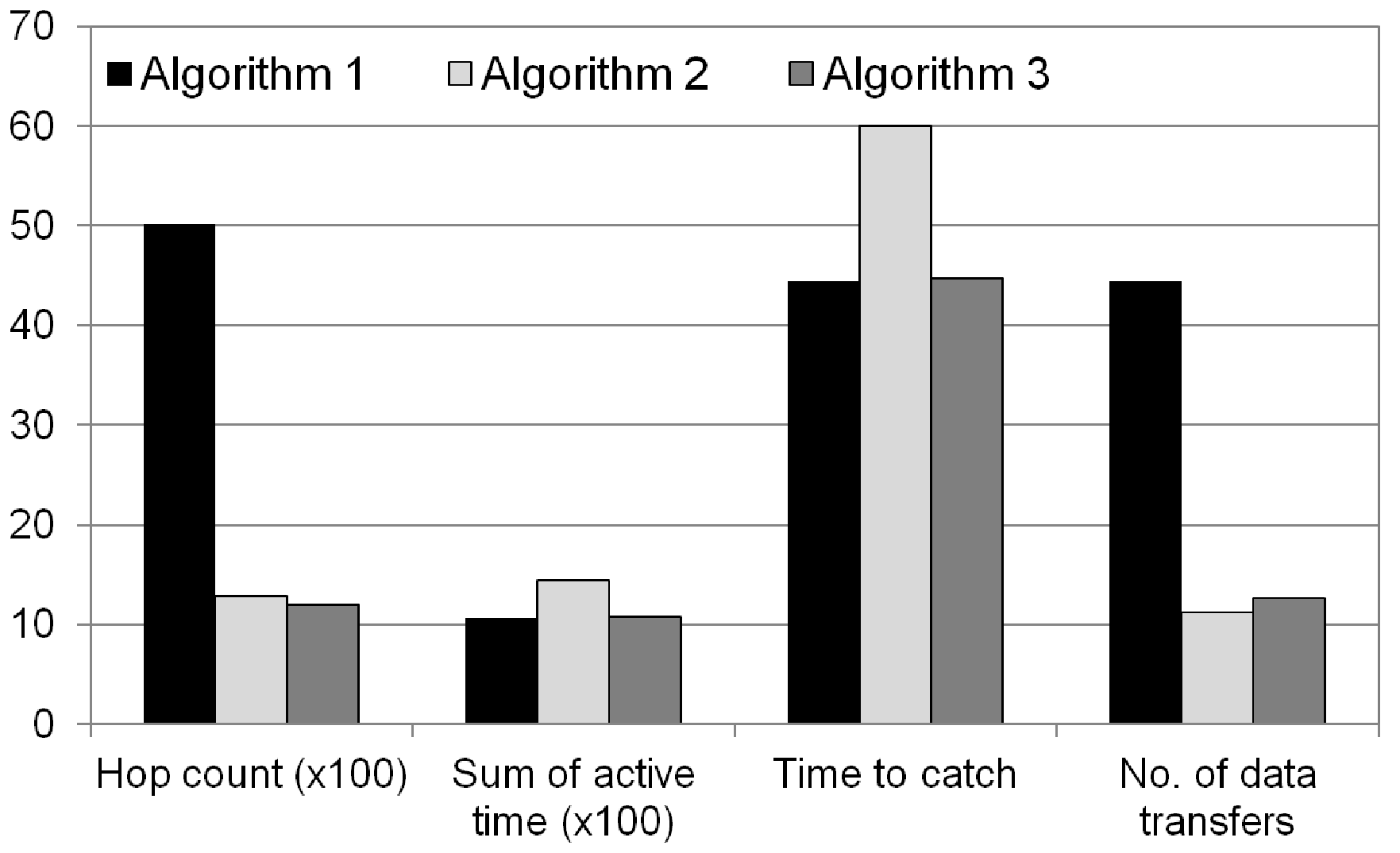}
\caption{Comparison of data collection costs and tracking performance}
\end{figure}

An example of a simulated target trajectory and resulting sink trajectories are presented in Fig. 6. The different trajectories of a sink were obtained by using the three examined algorithms. At the beginning of the simulation, target is located in segment with coordinates (66, 66) and sink is in segment (160, 160). For Algorithm 1 as well as for Algorithm 3 the target is caught after 62 time steps in segment (75, 12). When using Algorithm 2 the sink reaches target in 85-th time step at segment (105, 51). In this example, the hop counts for Algorithms 1, 2, and 3 are 7442, 1975, and 1440 respectively.

According to the presented results, it could be concluded that Algorithm 3, which is based on the proposed heuristic rules, enables a significant reduction of the data collection costs and ensures good performance of the target chasing application. The comparison of simulation results for Algorithms 1 and 3 shows that Algorithm 3 reduces hop count by about 75\% and increases time-to-catch by 1\% on average. 
%
%
\begin{figure}
\centering
\includegraphics [width=10.4cm] {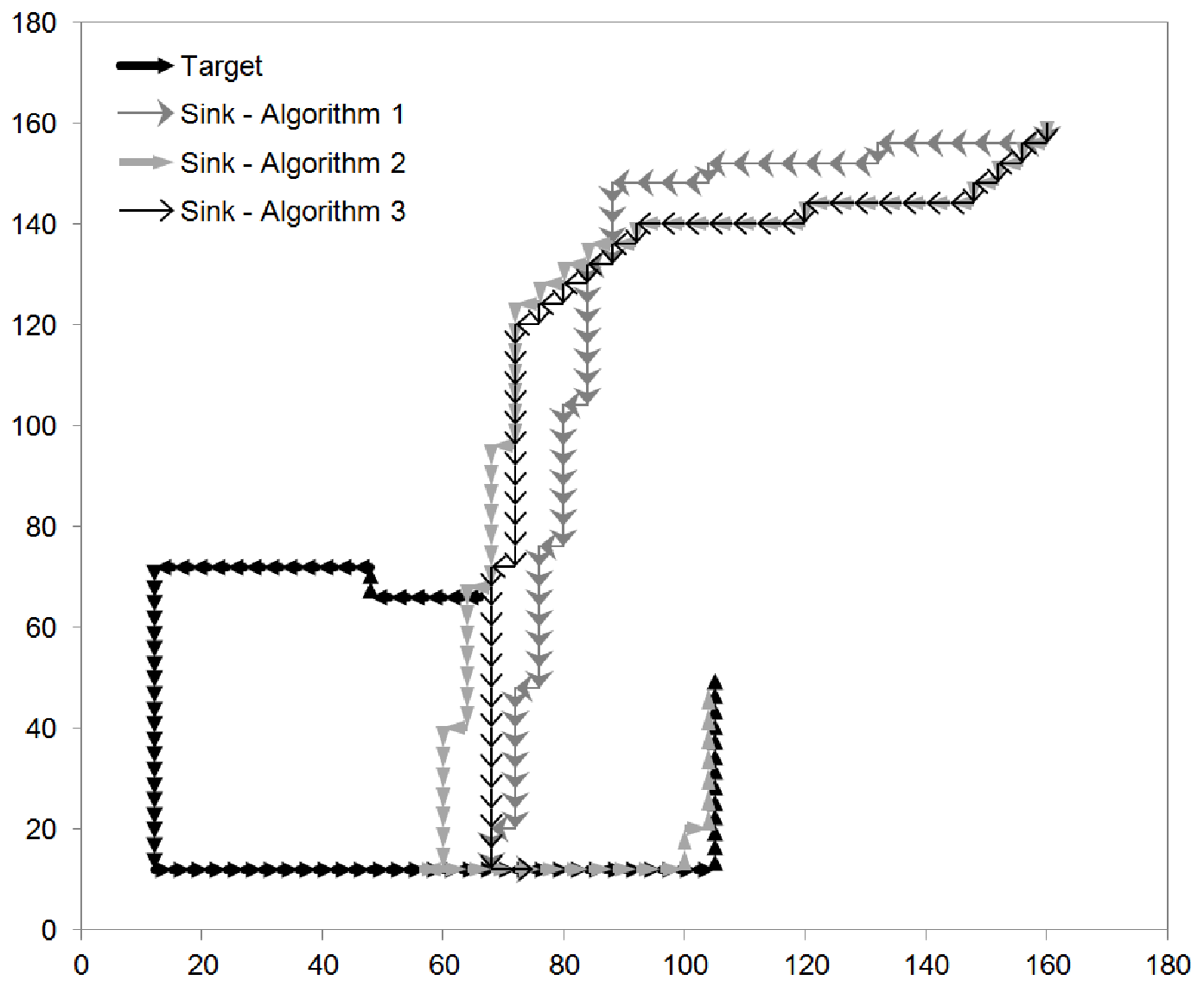}
\caption{Target trajectory and sink trajectories for compared algorithms}
\end{figure}
%
\section{Conclusion}
The WSN-based target chasing task requires dedicated methods for data collection. In order to optimize the utilization of WSN, the scope of the collected data has to be dynamically adjusted to the variable circumstances. In this paper, a tracking algorithm was proposed, which uses heuristic rules to decide when data transfers are necessary for achieving the chasing objectives. The heuristic rules are based on an observation that in some circumstances the sink does not need the precise information about target location to chase the target effectively.

Effectiveness of the proposed algorithm was compared against state-of-the-art methods. The experimental results show that the introduced heuristic rules enable substantial reduction in the data collection costs (hop count, sensor active time, and number of data transfers) with no significant increase in the amount of time necessary for mobile sink to catch the target. Future works will consider definition of the introduced heuristic rules and their uncertainty in terms of fuzzy sets.

%
%

\end{document}